\documentclass[amsmath,amssymb, aps, prl, twocolumn]{revtex4-2}
\usepackage{float}
\usepackage{graphicx}% Include figure files
\usepackage{dcolumn}% Align table columns on decimal point
\usepackage{bm}% bold math
\usepackage{hyperref}% add hypertext capabilities
\hypersetup{
    colorlinks=true,
    linkcolor=blue,
    filecolor=magenta,      
    urlcolor=cyan,
}
\usepackage{color}
\usepackage{fancyhdr}
\usepackage[normalem]{ulem}

\newcommand{\risp}[1]{\textcolor{black}{{#1}}}
\newcommand{\rizp}[1]{\textcolor{black}{{#1}}}

\begin{document}

\title{\risp{Testing the problem of time with cold atoms}}

\author{Giovanni Barontini}
\email{g.barontini@bham.ac.uk}
\affiliation{School of Physics and Astronomy, University of Birmingham, Edgbaston, Birmingham, B15 2TT, UK}

\date{\today}

\begin{abstract}
\rizp{We realize a cold-atom system to quantitatively test relational constructions of time. A well-isolated atomic Bose–Einstein condensate evolves in a conservative trap that is partitioned by a thin optical barrier into an observed and unobserved sector, with negligible dissipation on the experimental timescale. Motivated by relational-time approaches discussed in the Wheeler–DeWitt framework, we ask whether the dynamics of the observed sector can be ordered using only internal degrees of freedom. To this end, we construct an entropic time from an experimentally defined coarse-grained entropy, and demonstrate that it can robustly order the events in the observed sector across repeated cycles of expansion and recollapse. We finally derive an effective Schr\"odinger equation parameterized by this internal time and show that it is able to reproduce the measured evolution. These results establish a controlled experimental setting in which relational-time constructions can be quantitatively tested.}
\end{abstract}

\maketitle

Time is an outstanding problem in canonical quantum gravity.  The Wheeler–DeWitt (WDW) equation $\hat{H}\Psi=0$ admits no external parameter with which to sequence physical changes, in apparent contrast with our experience of time flowing \cite{PhysRev.160.1113}. Another fundamental problem is represented by the so-called arrow of time. All our theoretical frameworks, from Newtonian to quantum mechanics, relativity, and the above mentioned WDW equation, offer no built-in temporal orientation. {The only robust asymmetry in fundamental physics is the second law of thermodynamics, which pushes toward larger coarse-grained entropy. A standard explanation is the past hypothesis: the universe began in an exceptionally low entropy macrostate, see, e.g., \cite{albert2003time,zeh2007physical}. In a WDW description however, the total state is pure and the fine-grained entropy is conserved. At first sight, this appears incompatible with the observed growth of coarse-grained or reduced entropy.}

An array of strategies have been devised both for reinstating an effective temporal variable and enabling some form of entropy to flow within the WDW framework, see, e.g., \cite{PhysRevD.28.2960,PhysRevDD.31.1777, PhysRevDx.62.043518,Kuchar:1991qf,KieferSandhöfer+2022+543+559,Isham:1992ms,Connes_1994, PhysRevDwoo.27.2885}. Across these strategies, a consistent approach emerges: by partitioning the universe into subsystems, one can define some internal entropy that flows between one subsystem to another. In such models, one part of the universe can act as an entropy sink or source for another, even though the overall entropy remains constant. In this framework, the simplest approach is perhaps the one of minisuperspace models \cite{PhysRevD.28.2960,PhysRevDD.31.1777,PhysRevDf.50.2581}, in which strong symmetry restrictions such as spatial homogeneity or isotropy are imposed, so that the description of the universe is truncated to a finite number of degrees of freedom. In canonical minisuperspace models, what we call \emph{time} can re-appear as an \emph{emergent}, relational parameter when one dynamical variable is promoted to be the \emph{clock}, and all other variables are expressed in relation to it \cite{Halliwell:1989myn, Isham:1992ms,KieferSandhöfer+2022+543+559,Kuchar:1991qf,PhysRevDa.11.768,PhysRevD1.51.5600}. {Even in these simple models, the arrow of time remains an interesting issue, see e.g. \cite{Kuchar:1991qf,Isham:1992ms,KieferSandhöfer+2022+543+559,PhysRevD2.51.4145}, that can be addressed for example by tracing over environmental variables and accounting for decoherence. This yields effective mixed states whose reduced entropy can increase even though the global state remains pure \cite{Chataignier_2025,cespedes2025cosmologydecoherencesecondlaw}. An early prototype of this mechanism is Mott’s analysis of cloud chamber tracks, where classical records arise from entanglement with many atoms while the global state remains pure \cite{mott1929wave,Mott_1931}.} 

In recent years, cold atom platforms have evolved into quantitative quantum simulators for a variety of high-energy and cosmological models, including curved spacetime quantum field theory and lattice gauge dynamics, enabling laboratory access to questions traditionally reserved for cosmology and quantum gravity. For example, analogue black hole horizons in Bose–Einstein condensates have revealed spontaneous Hawking radiation \cite{steinhauer2016observation,munoz2019observation}, supersonically expanding ring condensates have emulated a Friedmann–Robertson–Walker universe \cite{PhysRevX.8.021021}, programmable Rydberg arrays and trapped ions have imaged the analogous of string breaking \cite{gonzalez2025observation,de2024observationstringbreakingdynamicsquantum}, and ultracold gases have observed bubble nucleation and Schwinger-like pair production during controlled false vacuum decay \cite{zenesini2024false,zhu2024probingfalsevacuumdecay}. 

\rizp{Motivated by the problem of time in the WDW framework, we realize a well isolated cold atom system with a time-independent Hamiltonian, and use it as an experimental analogue to probe relational time constructions. By using optical dipole potentials, we partition the system into a 'dark sector' and a 'bright sector', and we show that the effective Hamiltonian of the bright sector is structurally analogous to standard minisuperspace models. In this spirit, we identify one of the analogous fields as a clock, and from this we define an \emph{entropic time}. We show that this entropic time is a meaningful internal time variable over which it is possible to order the dynamics of the bright sector. We then derive an entropic time Schr\"odinger equation for the remaining degree of freedom, and demonstrate that it can be used to reproduce the experimental data. Our work effectively provides an experimentally grounded testbed for assessing internal-time constructions and entropy-based arrows of time in a controlled many-body setting.}  

\begin{figure}
\centering
\includegraphics[width=0.48\textwidth]{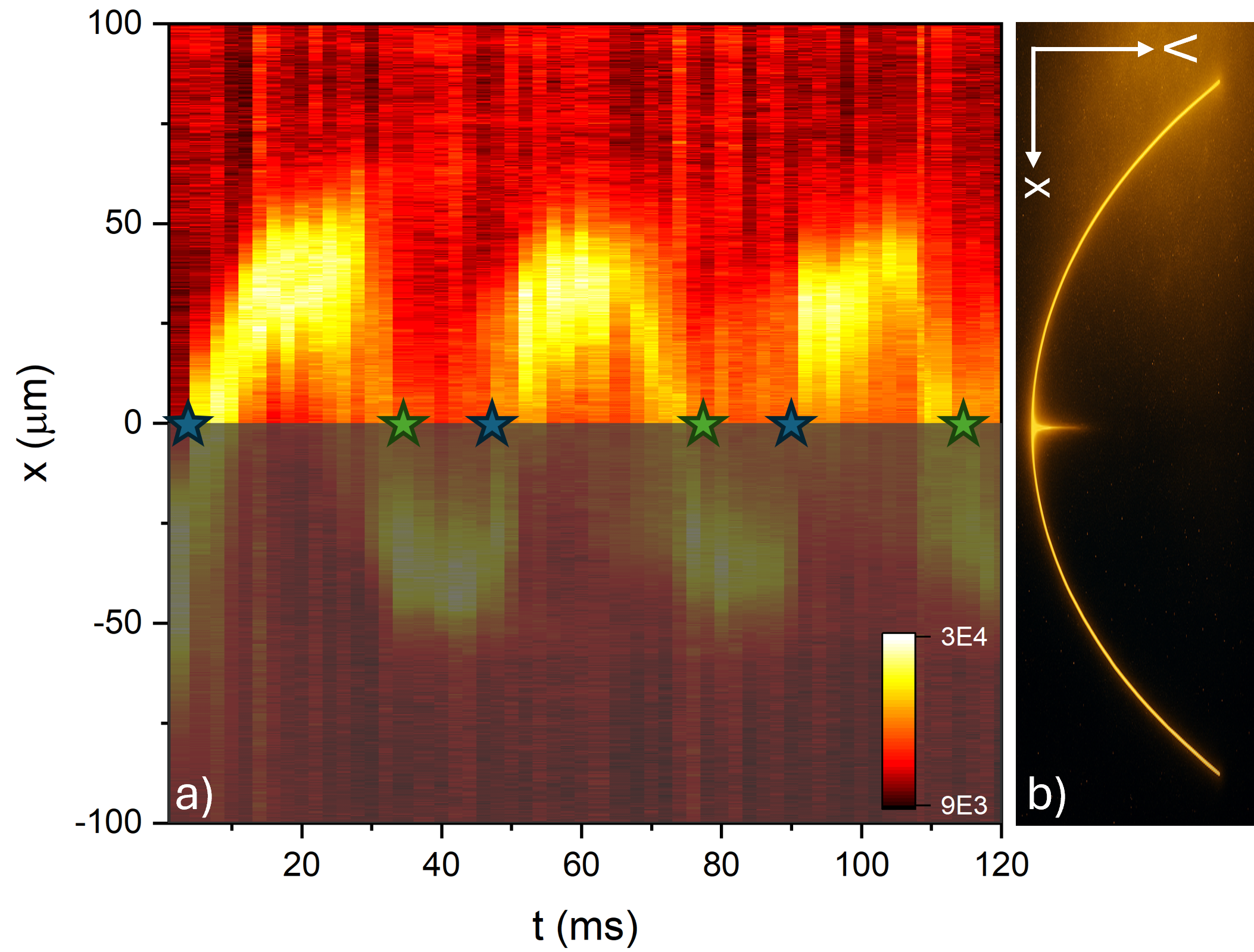}
\caption{{a) Experimental absorption images integrated along the $y$ axis showing the evolution of our \rizp{system} as a function of the external lab time. The colour scale indicates the integrated column density in arbitrary units. Our Bose-Einstein condensate evolves in a conservative harmonic trap that has a thin potential barrier at the bottom, as shown in panel b) \risp{using arbitrary units}. The barrier separates the 'dark' (unobserved) from the 'bright' (observed) sector. In the experiment these correspond to the bottom (shaded) and top half of panel a). Depending on the height of the barrier, the condensate is able to cross over from one sector to the other. The moment the atoms start to populate the bright sector corresponds to the 'big bang' (blue stars), while the moment it comes back to the dark sector corresponds to the 'big crunch' (green stars). \risp{The data shown in panel a) correspond to $V\simeq0$ and trapping frequencies $\simeq2\pi\times(25,70,70)$ Hz.}}}
\label{Fig1}
\end{figure}

Our typical experiment is shown in Fig. \ref{Fig1} and can be very easily described using the standard \emph{lab time}: a Bose-Einstein condensate of {$\simeq2.4\times10^4$ $^{87}$Rb atoms} oscillates back and forth along the $x$ direction in a conservative, radially symmetric, optical dipole trap. The trap is made by crossing a beam at 1070 nm and one at 1550 nm, resulting in final trapping frequencies of $\simeq2\pi\times(25,70,70)$ Hz. In the plane $x=0$, a thin potential barrier with a width of $\simeq 8$ $\mu$m is generated using a digital micromirror device. The light producing the potential barrier is at 675 nm. {Additional details are provided in \cite{SM}}. On the timescale of interest of this experiment ($\simeq100$ ms), the system does not experience any measurable dissipation or particle loss. \rizp{Because the system is well isolated during evolution, it is essentially a closed system with a time-independent Hamiltonian.} In analogy with the problem of time in WDW models, an issue then arises if one wants to describe the dynamics of this \rizp{isolated system} without using parameters that are external to it, such as the lab time that we have used so far. \rizp{Addressing this operationally in our controlled experiment can provide a quantitative benchmark for relational time constructions, and helps identify when an internal ordering parameter can replace the external one, thereby providing an operational analogue of the time-free descriptions sought in WDW models.}

\rizp{In what follows we consider our system as a 'mini-universe’, in the minimal sense of an effectively closed many-body system. As described above, this universe is separated into a 'dark' (unobserved) sector, on one side of the barrier, and a 'bright' (observed) sector, on the other side. This separation resembles those between “clock” and “rest” variables \cite{PhysRevDwoo.27.2885}, or “heavy” and “light” degrees of freedom in Born–Oppenheimer-type treatments \cite{kiefer2005semiclassical}.} The total Hamiltonian can thus be written as
\begin{equation}
\hat{H}=\hat{H}_{bright}+\hat{H}_{dark}+\hat{H}_{coupling}.
\end{equation}
\rizp{We then concentrate on the description of the bright sector only.} {Returning for a moment to the use of the external lab time, we can observe the typical dynamics of the bright sector in the absorption pictures of Fig. \ref{Fig1} (a). After a 'big bang' (blue stars), the \rizp{bright sector} grows until it reaches its maximum extension. It then contracts and finally collapses into the 'big crunch' (green stars).} The amount of atoms entering or escaping the bright sector is determined by the height of the barrier, i.e. the magnitude of $\hat{H}_{coupling}$. 

\rizp{We model the bright sector by an effective non-Hermitian Hamiltonian $\hat{H}_{bright}^{eff}=\hat{H}_{bright}-i\hbar\Gamma/2 $, where the last term accounts for the gain and loss of atoms across the barrier. This provides an effective gain-loss description of the bright sector due to its coupling to the unobserved dark sector, and underlies the entropy exchange used below to construct the internal time.} In the mean-field regime, $\hat{H}_{bright}$ is the usual Gross-Pitaevskii Hamiltonian, so that in spherical coordinates we can write \cite{FABREDELARIPELLE1983281}:
\begin{eqnarray}
&&\hat{H}_{bright}^{eff}=-\frac{\hbar^2}{2M}\nabla_X^2+\frac{1}{2}M\omega^2X^2 \nonumber \\
&&-\frac{\hbar^2}{2m}\left(\frac{\partial^2}{\partial R^2}+\frac{2}{R}\frac{\partial}{\partial R}\right)+\frac{1}{2}m\omega^2R^2 +\frac{g}{\Sigma(X)^3},  
\label{GPE}
\end{eqnarray}
where the \rizp{effective gain-loss contribution} has been absorbed in the  normalization condition $\int_{bright}|\psi|^2=N(X)$ \cite{PhysRevLett.110.035302}. Here $X$ is the $x$ coordinate of the center of mass (with good approximation, $Y=Z=0$ in our system), $R$ the \risp{hyperspherical} radial coordinate, $\Sigma$ the \risp{RMS} radius of the condensate, $M(X)=N(X)m$, with $N(X)$ the number of atoms and $m$ their mass, and $g$ the mean-field interaction. For simplicity, in the interaction term we have used the average density. This approximation, and the use of spherical instead of axial coordinates, greatly simplify the notation without affecting the physical content. \risp{Note that all quantities appearing in Eq. (\ref{GPE}) are defined on the bright sector only. In particular, $N(X)$ is the number of atoms currently in the bright sector, and $X$, $R$, and $\Sigma$ concern the bright-sector condensate. }

\rizp{At the level of the reduced degrees of freedom retained, Eq. (\ref{GPE}) is structurally similar to a WDW minisuperspace model, with $X$ playing the role of a uniform massive scalar field $\phi$, and $R$ the one of the scale factor $a$ \cite{KieferSandhöfer+2022+543+559}.} The term proportional to $\Sigma^{-3}$ acts as an effective 'dust' potential  \cite{wang2005dust}. {We note that the kinetic term in Eq. \ref{GPE} is positive-definite, whereas in canonical WDW equations the DeWitt metric induces an indefinite (hyperbolic) structure, with the scale factor typically carrying a negative kinetic term \cite{Halliwell:1989myn}. However, this sign is not invariant under field redefinitions, and what matters physically is the relational role of the chosen variables. Since in our case the internal time is defined via entropy flow rather than semiclassical propagation near a classical turning point, the positive-definite structure does not affect the phenomenology.} Above, we implicitly took $\phi$ as the clock field, through the relations $N=N(\phi)$ and $\Sigma=\Sigma(\phi)$. This is a rather common choice in minisuperspace models, but it comes with a caveat: in a recollapsing \rizp{bright sector} like ours, $\phi$ does not evolve monotonically. This means that $\phi$ cannot be a global time coordinate, because it is ambiguous concerning the direction of the evolution. {In recollapsing minisuperspace models the arrow aligns with increasing scale factor and reverses at the classical turning point; there is no single arrow spanning big bang to big crunch \cite{PhysRevD2.51.4145}. Several different strategies have been proposed to deal with the non-monotonicity of $\phi$ \cite{KieferSandhöfer+2022+543+559,mosta10.1063/1.532522}, and the relation between an emergent arrow of time and entropy has been developed in this context via decoherence and coarse graining \cite{Chataignier_2025,cespedes2025cosmologydecoherencesecondlaw}.}

Here we open a different pathway and, in analogy with what is done in stochastic systems \cite{pigoPhysRevLett.119.140604}, we define the entropic time for our \rizp{bright sector} as:
\begin{equation}
    \tau(\lambda)=\frac{\sigma}{k_B}\int_\lambda\frac{dS}{d\phi}|d\phi|,
    \label{eqtau}
\end{equation}
where $k_B$ is the Boltzmann's constant, $S$ the entropy \risp{in the bright sector}, $\sigma$ the (arbitrary) entropic time unit, and $\lambda$ defines the trajectory of $\phi$ \risp{within the bright sector}. This definition ensures that, as long as $dS$ and $d\phi$ have the same sign, the arrow of time does not change direction. Our experimental platform enables us to directly test this definition and verify that it provides us with a meaningful ordering of our data. {We note that Eq. \ref{eqtau} bears a formal resemblance to the semiclassical WKB time introduced in Born–Oppenheimer-type expansions of the WDW equation \cite{kiefer2005semiclassical},  with the difference that our construction uses the thermodynamic entropy rather than a Hamilton–Jacobi function.}

\begin{figure}
\centering
\includegraphics[width=0.44\textwidth]{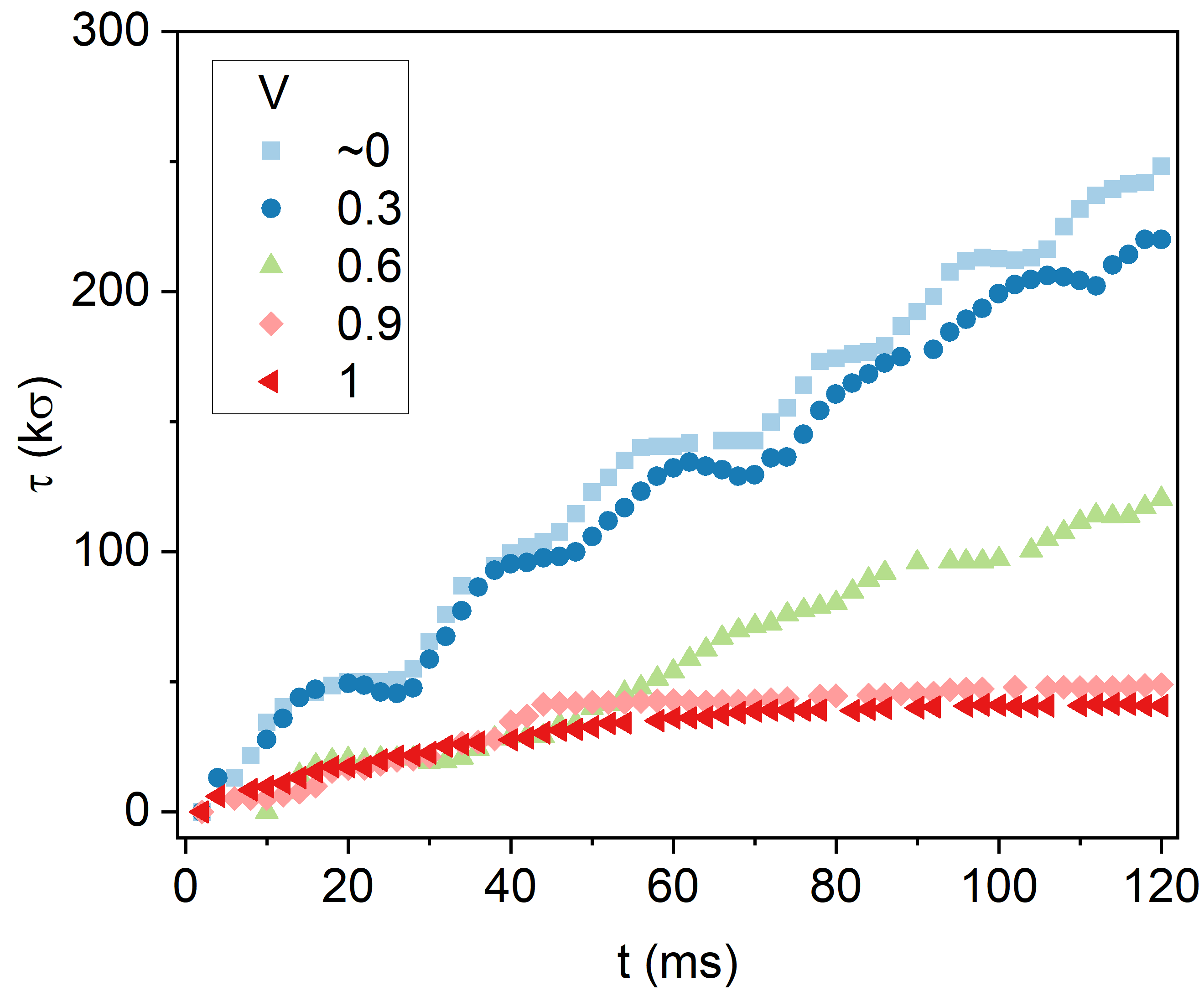}
\caption{{The solid symbols are the entropic internal time for the bright sector evaluated from experimental data using Eq. \ref{eqtau}, as a function of the external lab time. The experiment has been repeated for different values of the height of the potential barrier that separates the dark from the bright sector. The relative 1 $\sigma$ statistical uncertainty associated with each point (not shown) is approximately 5\%.}}
\label{Fig2}
\end{figure} 

In our experiment, we follow the dynamics in lab time for 120 ms, taking an absorption image every 2 ms, as shown in Fig. \ref{Fig1}. We repeat the measurement for different values of the height of the potential barrier, therefore \rizp{changing the exchange of entropy between the sectors}. For each image we measure $N$ and $\phi$ by calculating the integral and the center of mass of the density profiles, and the entropy per atom $s$ by utilizing the method of \cite{made22090956}, from which we derive $S=Ns$ \footnote{Since the entropy per particle 
$s$ remains of order unity throughout the evolution, the total entropy $S$ is effectively proportional to the number of atoms in the bright sector. Thus, entropy flow is directly linked to atom number dynamics.}. For each image we also measure the standard deviation $\Sigma$ of the density profile. Starting from the first big bang, we compute $\tau$ according to Eq. (\ref{eqtau}). For a direct comparison, we report in Fig. \ref{Fig2} the values of $\tau$ as a function of the external lab time, for different values of the height of the potential barrier $V=H_{coupling}/H_{coupling}^{max}$, with $H_{coupling}^{max}/k_B\simeq$ 255 nK. Crucially, $\tau$ grows monotonically almost everywhere. Its slope with respect to the lab time is set by the flow of entropy in the bright sector, so that the entropic time flows faster when entropy is \rizp{transferred in or out of the system}, and stops when no entropy is exchanged with the dark sector. For all our data set, we have verified that, within our errorbars, the entropy of the whole \rizp{mini-universe} (dark and bright sector) is constant.

\begin{figure}
\centering
\includegraphics[width=0.48\textwidth]{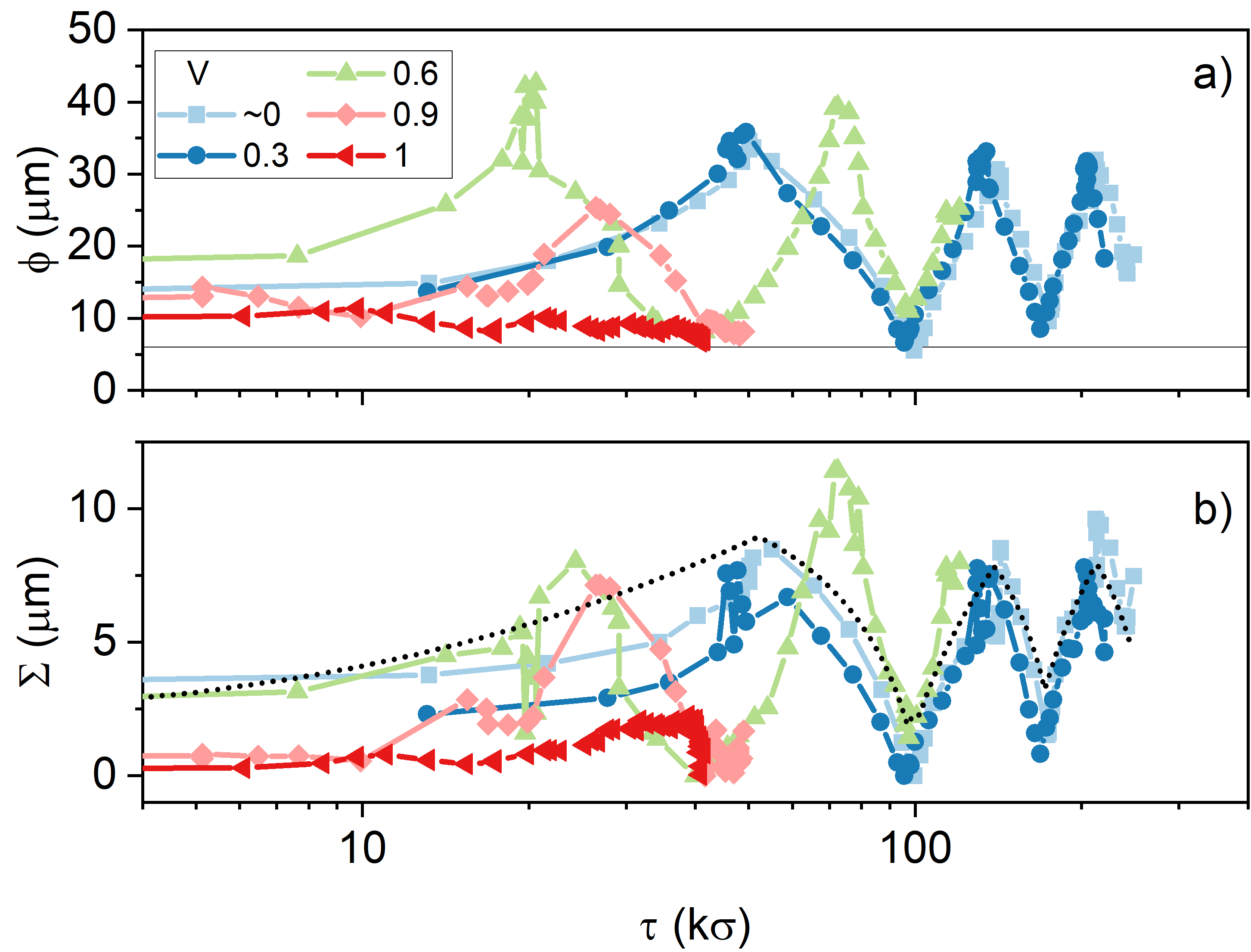}
\caption{a) {Experimentally} measured values of the analogue massive scalar field $\phi$ of the bright sector as a function of the entropic time for different values of the height of the barrier potential. As described in the text, in the experiment $\phi$ corresponds to the $X$ component of the center of mass of the condensate. The horizontal line approximately corresponds to the edge of the potential barrier (centered at $\phi$=0). b) {Experimentally} measured value of the width of the condensate {$\Sigma$} in the bright sector (corresponding to the size of the analogue universe) as a function of the entropic time, and for different values of $V$. The dotted curve is the result of the numerical simulations using Eq. (\ref{SE}) with $V\simeq0$. {In both panels, the relative 1 $\sigma$ statistical uncertainty associated with each point (not shown) is approximately 5\%.}}
\label{Fig3}
\end{figure} 

In Fig. \ref{Fig3} we report the measured values of $\phi$ and $\Sigma$ ordered with respect to the entropic time. With the exception of a few 'wiggles' where $dS$ and $d\phi$ change sign, caused by the coarse sampling in the clock field $\phi$, the ordering of the data broadly reflects the one in lab time. The spacing between the data points, which indicates the speed at which time is flowing, is however very different, as expected from what reported in Fig. \ref{Fig2}. For low values of $V$ we observe the cycling evolution of the \rizp{bright sector} from the big bang to the big crunch (as we do using the external lab time). For these settings the exchange of entropy between the dark and the bright sector is almost entirely reversible. In contrast with what observed in lab time, no entropic time elapses between a big crunch and the subsequent big bang, because no entropy is exchanged there. Note in Fig. \ref{Fig3} a) that, because the center of the potential barrier is at $\phi=0$, $\phi$ is bound by the thickness of the barrier, so that the bright sector never experiences the 'singularity' in the big bang or big crunch. For higher values of $V$ the exchange of entropy is progressively reduced and, as a result, the entropic time flows slower (although in lab time all traces have the same length). For V$\simeq$ 1 we reach the conditions for which the dynamics of the bright sector is no longer cyclic, but instead evolves towards its 'heat death', where the entropic time completely stops (corresponding to a stationary state in lab time).   

\begin{figure}
\centering
\includegraphics[width=0.48\textwidth]{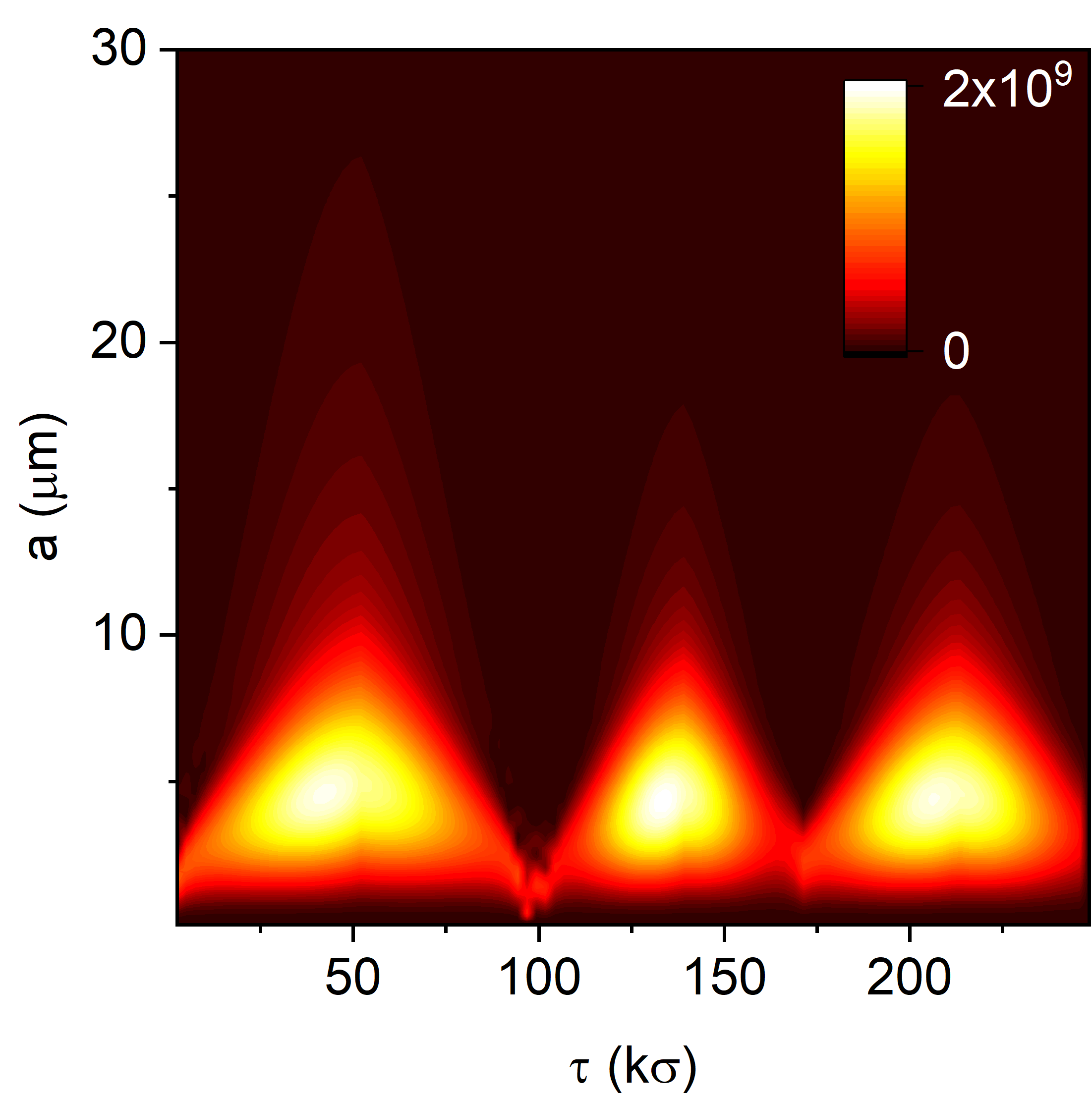}
\caption{Density probability distribution of the \rizp{bright sector} $N(\tau)|\psi(\tau,a)|^2$ as a function of the entropic time $\tau$ and the scale parameter $a$, obtained by numerically solving Eq. \ref{SE} using the experimental parameters of the data set with $V\simeq$ 0 shown in Fig. \ref{Fig2} and \ref{Fig3}.}
\label{Fig4}
\end{figure} 

From the test on our experimental data, the above definition of entropic time appears to be a meaningful choice for the internal time variable, providing a robust arrow that is directly linked to the entropic dynamics. The next step is to derive a (entropic) time-dependent Schr\"odinger equation for the wavefunction of the bright sector, starting from the (lab) time independent Hamiltonian of Eq. \ref{GPE}. To do so, we write the (lab) time independent Schr\"odinger equation:
\begin{equation}
    -\frac{\hbar^2\partial^2_\phi}{2M}\psi(\phi,a)+\left[\frac{1}{2}M\omega^2\phi^2+H_{geom}\right]\psi(\phi,a)=0,
 \label{eq4}   
\end{equation}
where $H_{geom}$ includes all the terms in the second row of Eq. \ref{GPE}. In addition, we make the approximation $M(\phi)=\alpha\phi$, which for our case is well justified. Using a Feshbach-Villars decomposition for $\psi$, \risp{which splits Eq. (\ref{eq4}) into two first-order equations with opposite signs of $\phi$-propagation, and keeping only the 'positive' onward-in-time solutions \cite{mosta10.1063/1.532522}}, and finally inserting the definition of $\tau$ as per Eq. (\ref{eqtau}), we obtain the non-local Schr\"odinger equation:
\begin{equation}
    i\hbar\partial_\tau\psi(\tau,a)={d_\tau\phi}\sqrt{\alpha^2\omega^2\phi^4+2\alpha\phi H_{geom}}\psi(\tau,a).
\end{equation}
The first term under the square root is $\approx N$ times the second term, therefore it is the dominant one when the system is away from the big bang or the big crunch ($\phi\neq0$). Performing a simple Taylor expansion we then obtain the entropic time Schr\"odinger equation:
\begin{equation}
    i\hbar\partial_\tau\psi(\tau,a)= \Phi(\tau)\psi(\tau,a)+\Lambda(\tau)H_{geom}\psi(\tau,a),
    \label{SE}
\end{equation}
where $\Phi=\alpha\omega\phi^2d_\tau\phi$ is a global phase and $\Lambda=({\partial_\phi S})^{-1}{k_B}/{\sigma\omega\phi}$. The $\Lambda$ factor in front of $H_{geom}$ is effectively an entropy dependent energy pump. Its time derivative controls whether energy flows into the $a$ degree of freedom ($\partial_\tau\Lambda>0$) or is sucked out ($\partial_\tau\Lambda<0$). \risp{When $\partial_\tau\Lambda$ is small but nonzero, Eq. (\ref{SE}) reduces to a Schr\"odinger equation with a slowly varying effective Hamiltonian. In this adiabatic regime the instantaneous eigenstates of $H_{geom}$ remain approximately stationary. Ordinary quantum mechanics thus applies locally, with $\Lambda(\tau)$
acting as a slowly varying energy scale, and corrections come at order $\partial_\tau\Lambda$. The strictly unitary Schr\"odinger dynamics is recovered only in the limit of zero entropy flow, $\partial_\tau\Lambda=0$, which is why Eq. (\ref{SE}) is more general than the ordinary one whenever the observed system is thermodynamically open to an unobserved one.} As expected, the equation is not well defined when there is no entropy -and therefore time- flow.

We can now use Eq. (\ref{SE}) to numerically reproduce our experimental observations, concentrating as an example on the case $V\simeq0$ (where the role of the entropic pump is most dominant). To do so we use a standard split-step-Fourier method on a $a$ grid of 8192 sites of total length 200 $\mu m$. The time step used is $d\tau=25$ $\sigma$, for a total duration of 250$\times10^3$ $\sigma$. From the experimental data we infer $\alpha\simeq5\times10^8$ m$^{-1}$kg, and the behavior of the entropy dependent pump $\Lambda$. As initial state we choose a simple Gaussian with width equal to the harmonic oscillator length. The results of our simulations are shown in Fig. \ref{Fig4}, where we report $N(\tau)|\psi(a,\tau)|^2$. As expected, for this configuration the dynamics is completely dominated by the behavior of the $\Lambda$ pump. We then fit the density profiles with a Gaussian function and we plot in Fig. \ref{Fig3} b) the values obtained for the standard deviation $\Sigma$ as a function of $\tau$ (dotted line), finding excellent agreement with our data.

\rizp{In summary, we have realized a cold atom analogue platform motivated by the WDW problem of time, in which the bright sector subsystem can be described with a reduced model structurally similar to an analogue minisuperspace model. We have provided a recipe for building an emergent internal time that accounts for the entropy exchange between sectors within the isolated system, and shown that it delivers a robust and meaningful monotonic arrowed ordering parameter over which our data can be sequentially ordered. We have then derived a Schr\"{o}dinger equation in entropic time whose numerical solutions are able to quantitatively reproduce the measured evolution.} Our entropic time approach bears conceptual similarities to the thermal time hypothesis \cite{Connes_1994}. However, our entropic time is constructed operationally from measurable entropy exchange between subsystems, rather than from the algebraic structure of observables. While the thermal time hypothesis applies to equilibrium states, our construction is inherently dynamical and experimentally accessible. It would be interesting to investigate whether the two notions coincide in certain limits, where entropic time could perhaps serve as a laboratory analogue of thermal time. Our work demonstrates that cold atom systems can function as a controlled environment to benchmark relational time constructions and arrows of time by direct and quantitative comparison with experimental data. Building on the wide set of tools available to engineer the terms in the Hamiltonian, cold atom systems could be a useful platform for quantitative studies of WDW-motivated relational-time scenarios and related analogue questions. A few concrete examples include: i) to help address the problem of possible multiple choices for the internal clock that may lead to a change in the canonical structure \cite{PhysRevDg.51.5600} by measuring relative shifts between different clocks; ii) to investigate the role of singularities and bouncing scenarios during the big bang/big crunch \cite{PhysRevLettx.86.5227,chandran2025expansion} by controlling the sign and strength of the interactions to determine whether the system experiences a true singularity or a quantum bounce; iii) to perform accurate tests of reversibility \cite{PhysRevD2.51.4145} through Loschmidt echo, iv) to engineer analogue black holes in the bright sector \cite{KieferSandhöfer+2022+543+559,zzxxPhysRevD.107.L121502} by using arbitrarily shaped attractive potentials; v) to realize Vilenkin-type tunnelling scenarios \cite{PhysRevDf.50.2581} by engineering coherent Josephson tunnelling between the sectors. The entropic time concept, as well as our experimental approach, could, in principle, be generalized to more complex models, such as midisuperspace-inspired settings or larger many-body analogues exploiting the potentially available 6$N$ degrees of freedom \cite{barbero2010quantization}. Future work could therefore explore whether our approach yields consistent arrows of time under more general dynamical conditions, also leveraging tools from quantum technology and information.

\paragraph*{Acknowledgements}
I acknowledge fruitful discussions with the Atomic Quantum Systems group at the University of Birmingham. I am grateful to Giulio Barontini for inspiring the core idea of this work, and to Vera Guarrera and Amita B. Deb for reading the manuscript and useful comments. 

\bibliography{main_bibl}

\end{document}